# Ultraviolet, Optical, and Near-IR Microwave Kinetic Inductance Detector Materials Developments

P. Szypryt, B. A. Mazin, B. Bumble, H. G. Leduc, and L. Baker

*Abstract*—We have fabricated 2024 pixel microwave kinetic inductance detector (MKID) arrays in the ultraviolet/optical/near-IR (UVOIR) regime that are currently in use in astronomical instruments. In order to make MKIDs desirable for novel instruments, larger arrays with nearly perfect yield need to be fabricated. As array size increases, however, the percent yield often decreases due to frequency collisions in the readout. The per-pixel performance must also be improved, namely the energy resolution. We are investigating ways to reduce frequency collisions and to improve the per pixel performance of our devices through new superconducting material systems and fabrication techniques. There are two main routes that we are currently exploring. First, we are attempting to create more uniform titanium nitride films through the use of atomic layer deposition rather than the more traditional sputtering method. In addition, we are experimenting with completely new material systems for MKIDs, such as platinum silicide.

*Index Terms*—Infrared imaging, optical imaging, superconducting device fabrication, superconducting microwave devices, superconducting resonators

## I. Introduction

Microwave Kinetic Inductance Detectors (MKIDs) [1], [2] are a new type of superconducting technology capable of measuring the arrival times and energies of individual photons. MKIDs work using the principle of the kinetic inductance effect [3]. Energy can be stored in the supercurrent (or flow of Cooper pairs) of a superconductor. In order to reverse the direction of the supercurrent, energy must be removed from the superconductor, resulting in an additional kinetic inductance term. If a superconducting material is patterned into a resonator, light hitting the resonator will momentarily increase the kinetic inductance of the superconductor, thereby decreasing the frequency of resonance. The magnitude of this response is closely related to the energy of the incident photon. An array of these resonators can be read out using a single microwave feedline using a frequency domain multiplexing scheme [4]. As with traditional charge-coupled devices (CCDs), the energy of the incident photon must be above the bandgap energy in order for the photon to be absorbed. Superconductors have bandgap energies roughly 1000 times lower than that of silicon, allowing for the detection of much lower energy photons.

MKIDs are ideal in astronomy for observations of time-varying objects and those in which spectral information is important. Some examples of objects observed with ultraviolet/optical/infrared (UVOIR) MKIDs are ultra-compact binaries, pulsars, and galaxies. UVOIR MKIDs have been proven as successful astronomical detectors through the publication of the first two astronomy papers using MKIDs at any wavelength. These involved observations of the 33 millisecond spin period Crab Pulsar [5] and the 28 minute orbital period AM CVn system, SDSS J0926+3624 [6]. Observations were done using our MKID instrument, the Array Camera for Optical to Near-IR Spectrophotometry (ARCONS) [7]. In the future, MKIDs will be used for speckle nulling in two funded exoplanet imagers, the Dark-speckle Near-IR Energy-resolved Superconducting Spectrophotometer (DARKNESS) at the Palomar observatory and the MKID Exoplanet Camera (MEC) on the Subaru Telescope. Observations using ARCONS are also funded to continue.

Although MKIDs have begun producing results in astronomy, there is much room for improvement. Future instruments require much larger MKID arrays. From a fabrication standpoint, scaling up MKID arrays is straightforward. As the array size goes up, however, the percent pixel yield typically goes down. Non-uniformities in superconducting critical temperature across a device cause resonators to shift away from their intended resonant frequencies, resulting in resonator collisions in frequency space. Colliding resonators cannot be distinguished and therefore cannot be read out properly, reducing the usable pixel count. The per-pixel performance of MKIDs also needs to be improved. Improving the energy resolution is the main priority, followed by quantum efficiency. The energy resolution is mostly limited by two-level system (TLS) [8] and amplifier noise, whereas the quantum efficiency depends mostly on choice of material. There are fabrication processes used in similar low temperature detectors, such as development of optical cavities in transition-edge sensors [9], which have been shown to increase quantum efficiency. These methods, however, typically increase the TLS noise in the detectors resulting in lower energy resolution. We are investigating new material systems and fabrication techniques to address the most pressing issues of energy resolution and critical temperature uniformity. Current work is going into developing MKIDs using thin films of atomic layer deposition

This work was supported by a NASA Space Technology Research Fellowship.
P. Szypryt is with the Department of Physics, University of California, Santa Barbara, CA 93106 USA, e-mail: pszypryt@physics.ucsb.edu.
B.A. Mazin is with the Department of Physics, University of California, Santa Barbara, CA 93106 USA, e-mail: bmazin@physics.ucsb.edu.
B. Bumble, H. G. Leduc, and L. Baker are with NASA Jet Propulsion Laboratory, Pasadena, CA 91109 USA.

(ALD) titanium nitride and platinum silicide.

## II. ATOMIC LAYER DEPOSITION TITANIUM NITRIDE

The current standard superconductor used in UVOIR MKID fabrication is sputtered Ti-N [10]. Ti-N is an ideal MKID superconductor due to its high kinetic inductance fraction, which leads to a large resonator responsivity due to incident photons. A higher responsivity leads to a more accurate determination of the photon energy. Sputtered Ti-N films, however, suffer from non-uniformities in superconducting critical temperature across a wafer, which greatly reduces percent pixel yield. The critical temperature is tuned by controlling the stoichiometry in the Ti-N films, and deviations in the nitrogen flow rate during sputtering can cause the stoichiometry of the Ti-N film to vary across a wafer. Due to the high kinetic inductance fraction, slight deviations in critical temperature create fairly large differences in the actual resonator frequencies from their expected design frequencies. There have been multiple schemes developed to suppress these variations, such as stacking multilayers of stoichiometric Ti-N and pure Ti [11]. Atomic layer deposition of Ti-N is another proposed method to increase the uniformity of sub-stoichiometric Ti-N films, but much of this work is still in a very early phase. A previous study of the microwave properties of ALD TiN films was performed in Ref. [12].

We used a Beneq TFS 200 ALD system to grow thin films of Ti-N. The precursors used were $TiCl_4$ and $NH_3$. The precursor flow rates were the first two parameters that were varied in attempts to obtain a 1 K $T_C$ film. The ALD process temperature was the third parameter that was varied. No plasma power was used, and the reaction energy was supplied completely by the thermal energy. A list of initial processing parameters and results is shown in Table I.

Our initial tests showed that the critical temperature roughly decreased with processing temperature, when processed between 460-507 °C. It should be noted that there was evidence of chemical vapor deposition (CVD) in addition to ALD for the 507 and 495 °C processing temperature samples, indicated by a laterally non-uniform film. This is the likely cause of the broad low temperature superconducting transitions in these two films. Another important result was measuring high quality factors for resonators structured out of these films. These internal quality factors were at the upper limits of our measurement technique, which is limited by our relatively low coupling quality factors. Future work will involve finding ALD processing conditions that create a 1 K $T_C$ Ti-N film and characterizing the uniformity in $T_C$ across a device.

## III. PLATINUM SILICIDE

Platinum silicide was also explored as a possible replacement for titanium nitride as a MKID superconductor. Bulk stoichiometric PtSi has a $T_C$ of 1 K, but the $T_C$ is suppressed for films below 50 nm [13]. This results in a $T_C$ in the desired range without having to alter the stoichiometry and risk creating non-uniformities in the array, as in the Ti-N case. PtSi is a fairly common material in semiconductor processing, and the room temperature properties are quite well characterized [14]. PtSi also has a high room temperature resistivity similar to that of sputtered TiN, indicative of a high kinetic inductance fraction. Most importantly, it is fairly simple to get PtSi into the desired stoichiometric state, as this is also the thermally stable state. A layer of platinum on a silicon wafer can be easily annealed into a PtSi film.

To begin the deposition of our initial PtSi films, we cleaned a (100) high resistivity silicon wafer. We used nanostrip followed by HF to remove the native oxide. The wafer was immediately brought into a CHA Industries SEC600 e-beam evaporator, where a 20 nm film of platinum was grown. The film was then brought into UHV and annealed at 500 °C for 20 minutes. This acts to create a PtSi layer that is roughly double the thickness of the initial platinum layer. The PtSi layer is then patterned into MKID test devices.

The initial samples showed $T_C$ of roughly 800 mK, which is within the optimal range of our cryogenic system. Photon events created quasiparticles with ~20 microsecond lifetimes, measured by observing the duration of a phase difference caused by a photon event. The measured energy resolution was 8 at 400 nm, which is equal to the energy resolution of our sputtered Ti-N films [7]. Energy resolution is measured by looking at the phase response of resonators due to photon events from lasers of precisely known wavelengths. In our sputtered Ti-N films, the measured quantum efficiency is 70% at 400nm and 25% at 1 micron. Preliminary count rate measurements in PtSi detectors indicate a similar quantum efficiency, and more precise quantum efficiency

TABLE I
INITIAL ALD TITANIUM NITRIDE RESULTS

| Process Temperature (°C) | Number of Cycles | Dose/Purge Times TiCl4 (ms) | Dose/Purge Times NH$_3$ (ms) | Critical Temperature (K) | Internal Quality Factor |
|---|---|---|---|---|---|
| 507 | 2000 | 100/1500 | 250/3000 | 0.250-0.800 | - |
| 495 | 1000 | 100/1500 | 200/4000 | 0.150-0.500 | - |
| 490 | 1000 | 100/1500 | 100/3000 | 1.547-1.558 | - |
| 490 | 1000 | 50/1000 | 200/4000 | 2.121-2.130 | - |
| 460 | 1000 | 100/1500 | 200/4000 | 3.2-3.8 | 200,000-300,000 |
| 460 | 2000 | 100/1500 | 100/3000 | 2.520-3.077 | 100,000-200,000 |
| 460 | 2000 | 50/800 | 100/2000 | 2.922-3.242 | >500,000 |

Initial results of thin TiN film depositions using ALD. Critical temperature decreased with increasing process temperature, but was largely uncontrolled. High resonator internal quality factors were measured, and these measurements were limited by the fairly low coupling quality factors of ~30,000. Not all samples were patterned. Only patterned samples had measured internal quality factors.

measurements will be made once the PtSi film thickness has been fully tuned for sensitivity. The quality factors were in the range of 10,000 to 30,000, which is much lower than what is observed in sputtered Ti-N films (Ref. [10] saw internal quality factors of substoichiometric Ti-N films of $5 \times 10^6$). Increasing this quality factor would make PtSi competitive as a replacement for Ti-N.

There are various methods that we are employing to try to increase the quality factors of our PtSi films. One method we investigated was attempting to alter the crystal phases of the PtSi films. The hope was that some crystal orientations were better than others for quality factor, an effect exhibited in our Ti-N films. We found that one way to alter crystal structure was vary the PtSi film thickness, as can be seen in the X-ray diffraction pattern in Fig. 1. Unfortunately we were unable to correlate the crystal phase differences to any significant increases in quality factor. Future work in altering the PtSi crystal structure will likely involve depositing platinum on different crystal orientation silicon substrates.

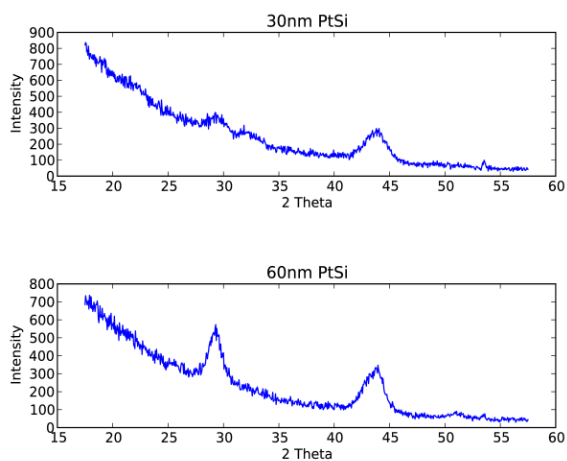

Fig. 1. X-ray diffraction patterns for PtSi films of 30 and 60 nm film thicknesses. The peak at ~29 degrees corresponds to the (101) orientation, whereas the peak at ~43.5 degrees corresponds to the (121) orientation. In our initial depositions, decreasing the film thickness acted to suppress the (101) crystal phase.

Another likely source for the low quality factors could be excess platinum diffusing into the silicon. Fig. 2 shows secondary ion mass spectroscopy measurements of a 40nm PtSi film grown on a silicon substrate. Instead of a sharp PtSi-Si interface, a gradual decrease in the platinum to silicon ratio for ~20nm is observed. For this reason, we attempted to grow a PtSi film on a sapphire substrate. To do this, we cleaned a sapphire wafer and deposited 25 nm of platinum via e-beam evaporation. Afterwards, the wafer was placed in an ICP PECVD system. The chamber pressure was held at 50 mTorr and temperature at 350 C, which was at the limits of the system. 30 sccm of $SiH_4$ was flowed through the chamber for upwards of two hours. Unfortunately, no appreciable PtSi layer was grown. It was apparent that a PECVD system with higher maximum gas pressures and temperatures would be required in order to match conditions of previous work with successful PtSi formation [15]. In the future, we will sputter platinum and silicon onto a sapphire substrate, and continue by annealing the sample as in the PtSi on silicon case.

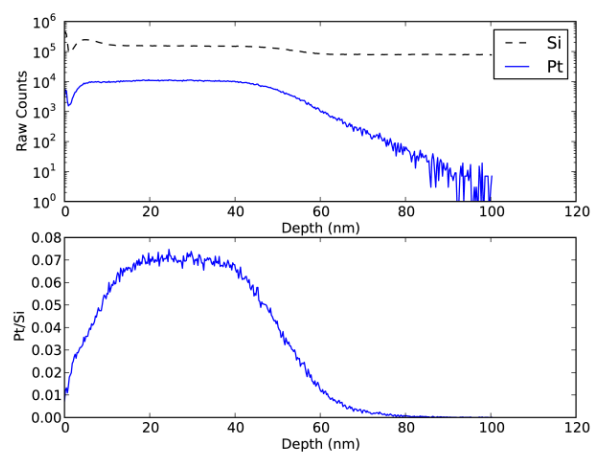

Fig. 2. Secondary ion mass spectroscopy measurements of 40 nm PtSi film on silicon. Instead of a sharp decrease in the Pt/Si ratio at 40 nm, there is a gradual decline of almost 20 nm. This excess diffusion of platinum into silicon could be the cause of the low quality factors and is the major motivation for using a sapphire substrate.

IV. CONCLUSIONS

We investigated ALD Ti-N and PtSi as possible materials for use as MKID superconductors. Early tests of ALD Ti-N growth showed promising internal quality factors, but more work needs to be done in tuning the superconducting critical temperature to 1 K. In addition to higher quality factors, this process is expected to produce more uniform films. Once a 1 K $T_C$ film can be repetitively deposited, we will perform extensive uniformity tests. For PtSi, producing films of the desired $T_C$ was a fairly straightforward process. The quality factors of these films, however, were quite low, and future work will go primarily towards addressing this issue. The most promising solution is creating a process for growing a PtSi film on a sapphire substrate. Sputtering platinum and silicon on a sapphire substrate, and then annealing in-situ, will likely be the next step in this work.

REFERENCES

[1] P. K. Day, H. G. Leduc, B. A. Mazin, A. Vayonakis, J. Zmuidzinas, "A broadband superconducting detector suitable for use in large arrays," *Nature*, vol. 425, pp. 817-821, 2003.
[2] B. A. Mazin *et al.*, "A superconducting focal plane array for ultraviolet, optical, and near-infrared astrophysics," *Optics Express*, vol. 20, no. 2, pp. 1503-1511, 2012.
[3] D. C. Mattis and J. Bardeen, "Theory of the Anomalous Skin Effect in Normal and Superconducting Metals," *Phy. Rev.*, vol. 111, no. 412.
[4] S. McHugh *et al.*, "A readout for large arrays of Microwave Kinetic Inductance Detectors," *Rev. Sci. Instrum.*, vol. 83, no. 044702, 2012.
[5] M. J. Strader *et al.*, "Excess Optical Enhancement Observed with ARCONS for Early Crab Giant Pulses," *The Astrophysical Journal Letters*, vol. 779, no. L12, 2013.
[6] P. Szypryt *et al.*, "Direct Detection of SDSS J0926+3624 Orbital Expansion with ARCONS," *Monthly Notices of the Royal Astronomical Society*, vol. 439, no. 3, Apr. 2014.
[7] B. A. Mazin *et al.*, "ARCONS: A 2024 Pixel Optical through Near-IR Cryogenic Imaging Spectrophotometer," *Publications of the Astronomical Society of the Pacific*, vol. 125, no. 933, pp. 1348-1361, Nov. 2013.

[8] J. Gao *et al.*, "Experimental evidence for a surface distribution of two-level systems in superconducting lithographed microwave resonators," *Appl. Phys. Lett.*, vol. 92, no. 15, 2008.
[9] A. E. Lita, B. Calkins, L. A. Pellochoud, A. J. Miller, and S. Nam, "High-Efficiency Photon-Number-Resolving Detectors based on Hafnium Transition-Edge Sensors," *AIP Conference Proceedings*, vol. 1185, pp. 351-354, 2009.
[10] H. G. Leduc *et al.*, "Titanium nitride films for ultrasensitive microresonator detectors," *Appl. Phys. Lett.*, vol. 97, no. 102509, 2010.
[11] M. R. Vissers *et al.*, "Proximity-coupled Ti/TiN multilayers for use in kinetic inductance detectors," *Appl. Phys. Lett.,* vol. 102, no. 232603, 2013.
[12] P. C. J. J. Coumou *et al*., "Microwave Properties of Superconducting Atomic-Layer Deposited TiN Films," *IEEE Transactions on Applied Superconductivity*, vol. 23, no. 3, pp. 7500404, Jun. 2013.
[13] K. Oto, S. Takaoka, K. Murase, and S. Ishida, "Superconductivity in PtSi ultrathin films," *J. Appl. Phys*., vol. 76, no. 5339, 1994.
[14] H. Bentmann, A. A. Demkov, R. Gregory, and S. Zollner, "Electronic, optical, and surface properties of PtSi thin films," *Phys. Rev. B*, vol. 78, no. 205302, 2008.
[15] Y. Takahashi, H. Ishii, and J. Murota, "New platinum silicide formation method using reaction between platinum and silane," *J. Appl. Phys*., vol. 58, no. 3190, 1985.